\documentclass[sn-vancouver,iicol]{sn-jnl}

\theoremstyle{thmstyleone}%
\newtheorem{theorem}{Theorem}
\theoremstyle{thmstyletwo}%

\theoremstyle{thmstylethree}%

\usepackage{multirow}
\usepackage[strings]{underscore}
\usepackage{physics,amsmath}
\usepackage{booktabs}
\newtheorem{hyp}{Hypothesis}
\usepackage{caption}

\begin{document}

\title[Region-Based Evidential Deep Learning]{Region-Based Evidential Deep Learning to Quantify Uncertainty and Improve Robustness of Brain Tumor Segmentation}


\author[1,2]{\fnm{Hao} \sur{Li}}\email{hao.li19@imperial.ac.uk}

\author[1]{\fnm{Yang} \sur{Nan}}\email{y.nan20@imperial.ac.uk}


\author[3,4]{\fnm{Javier} \sur{Del Ser}}\email{javier.delser@tecnalia.com}

\author*[1,5]{\fnm{Guang} \sur{Yang}}\email{g.yang@imperial.ac.uk}


\affil[1]{\orgdiv{National Heart and Lung Institute, Faculty of Medicine}, \orgname{Imperial College London}, \city{London}, \country{UK}}

\affil[2]{\orgdiv{Department of Bioengineering, Faculty of Engineering}, \orgname{Imperial College London}, \city{London}, \country{UK}}

\affil[3]{\orgname{TECNALIA, Basque Research \& Technology Alliance (BRTA)}, \city{Derio}, \country{Spain}}

\affil[4]{\orgname{University of the Basque Country (UPV/EHU)}, \city{Bilbao}, \country{Spain}}

\affil[5]{\orgname{Royal Brompton Hospital}, \city{London}, \country{UK}}


\abstract Despite recent advances in the accuracy of brain tumor segmentation, the results still suffer from low reliability and robustness. Uncertainty estimation is an efficient solution to this problem, as it provides a measure of confidence in the segmentation results. The current uncertainty estimation methods based on quantile regression, Bayesian neural network, ensemble, and Monte Carlo dropout are limited by their high computational cost and inconsistency. In order to overcome these challenges, Evidential Deep Learning (EDL) was developed in recent work but primarily for natural image classification. In this paper, we proposed a region-based EDL segmentation framework that can generate reliable uncertainty maps and robust segmentation results. We used the Theory of Evidence to interpret the output of a neural network as evidence values gathered from input features. Following Subjective Logic, evidence was parameterized as a Dirichlet distribution, and predicted probabilities were treated as subjective opinions. To evaluate the performance of our model on segmentation and uncertainty estimation, we conducted quantitative and qualitative experiments on the BraTS 2020 dataset. The results demonstrated the top performance of the proposed method in quantifying segmentation uncertainty and robustly segmenting tumors. Furthermore, our proposed new framework maintained the advantages of low computational cost and easy implementation and showed the potential for clinical application.

\maketitle

\section{Introduction}\label{sec1}

Automated brain tumor segmentation promises to provide more reliable measurements for cancer diagnosis and assessment, establishing new possibilities for high-throughput analysis \cite{jungoAnalyzingQualityChallenges2020}. Segmentation enables clinicians to determine tumor location, extent, and subtype. Additionally, brain tumor segmentation on longitudinal MRI scans can facilitate monitoring tumor growth or shrinkage. In current clinical practice, accurate segmentation of brain tumor regions is usually done manually by experienced radiologists, which is time-consuming and labor-intensive. Furthermore, manual labeling of results may involve human bias, as they rely on the physician's experience and subjective decision-making. On the other hand, automated segmentation techniques can reduce labor and human bias to provide efficient, objective, and reproducible results for tumor diagnosis and monitoring.

The performance of automated brain tumor segmentation methods has grown rapidly over the past few years. This development is due to the growth of annotated datasets and the advent of deep learning models that can leverage large amounts of data \cite{muhammadDeepLearningMultigrade2021a}. Most methods are based on fully convolutional neural networks (FCN) \cite{longFullyConvolutionalNetworks2015}, like U-Net \cite{ronnebergerUNetConvolutionalNetworks2015} and its variants \cite{dongAutomaticBrainTumor2017} for improving the performance of brain tumor segmentation. Recently, Transformers and self-attention, originating from natural language processing (NLP), have also been applied to medical image segmentation \cite{chenTransUNetTransformersMake2021}.

Although the segmentation results of deep neural networks are reported to be close to or comparable to human performance \cite{bakasIdentifyingBestMachine2019}, their robustness levels are low, and concerns remain with their clinical acceptability \cite{mullerRobustnessBrainTumor2020}. Possible reasons include the large variability in imaging properties such as artifacts and magnetic field strength, as well as the inherent heterogeneity of brain tumors, which are beyond the training dataset. Furthermore, human bias in dataset annotations can cause models also to inherit this bias. One possible direction to alleviate the reliability problem of deep neural networks is to use uncertainty estimation. The uncertainty reflects how confident the network predicts the class labels. Confidence studies can help identify areas of data dominated by lack of annotations (epistemic uncertainty) or noisy annotations (aleatoric uncertainty). Additional information from uncertainty estimation can be used to quantify segmentation performance or as a post-processing step to correct automatic segmentation. Clinically, uncertainty estimates can feed back potential error regions to guide or automate corrections or be used for patient-level segmentation failure detection \cite{jungoAnalyzingQualityChallenges2020}. Therefore, reliably quantifying the uncertainty of segmentation performance is critical in clinical applications. 

Popular methods for quantifying uncertainty in neural networks include quantile regression (QR) \cite{dasQuantileRegression2019}, Bayesian neural network (BNN) \cite{hintonKeepingNeuralNetworks1993,mackayPracticalBayesianFramework1992,hernandez-lobatoProbabilisticBackpropagationScalable2015}, ensemble-based \cite{nairExploringUncertaintyMeasures2020}, dropout-based \cite{galDropoutBayesianApproximation2016,lakshminarayananSimpleScalablePredictive2017,kendallBayesianSegNetModel2017}, and evidential deep learning (EDL) \cite{sensoyEvidentialDeepLearning2018,tsiligkaridisInformationAwareMaxnorm2021, tongEvidentialFullyConvolutional2021}. Simply interpreting the confidence scores of softmax/sigmoid outputs as event probabilities in a categorical distribution can lead to overconfident wrong prediction \cite{guoCalibrationModernNeural2017,mehrtashConfidenceCalibrationPredictive2020}. The classical BNN aims to capture uncertainty by learning the weight distribution of the network and approximates the integral of parameters by variational inference or Laplace approximation to estimate the posterior prediction distribution \cite{tsiligkaridisInformationAwareMaxnorm2021}. However, most BNNs are challenging to implement and train since model parameters have to be explicitly modeled as random variables \cite{gawlikowskiSurveyUncertaintyDeep2022}, which lack scalability in both architecture and data size \cite{hernandez-lobatoProbabilisticBackpropagationScalable2015}. Hence, subsequent approaches focused on being able to reuse the training pipeline and maintain scalability while providing reasonable uncertainty estimates. To this end, more intuitive and simple methods, such as learning an ensemble of deterministic networks \cite{lakshminarayananSimpleScalablePredictive2017,mehrtashConfidenceCalibrationPredictive2020} and introducing Monte Carlo dropout \cite{nairExploringUncertaintyMeasures2020} are proposed for brain tumor segmentation. On the downside, ensemble-based methods need to train multiple models from scratch, which is computationally expensive, and the introduction of dropouts results in inconsistent outputs \cite{kohlProbabilisticUNetSegmentation2019}. 

On the other hand, EDL has been gradually developed in recent studies, demonstrating more promising and reliable performance in uncertainty estimation. Based on the Dempster-Shafer Evidence Theory (DST) \cite{dempsterGeneralizationBayesianInference2008}, EDL uses the Dirichlet distribution to model the categorical distribution of the output given the input to the network. This class of methods produces closed-form prediction distributions and outperforms BNNs in adversarial queries and out-of-distribution uncertainty quantification \cite{tsiligkaridisInformationAwareMaxnorm2021}. Compared to ensemble-based and dropout-based methods, EDL showed more robust results with lower computational costs \cite{zouTBraTSTrustedBrain2022}. However, most of the recent works focus on the natural image classification and segmentation problem, making the application of EDL in medical image segmentation to be further studied.

In this paper, we propose a region-based EDL network for accurate brain tumor segmentation. The network learned about classification distribution by minimizing region-based prediction error under the Dirichlet prior distribution. This enabled the proposed network to provide accurate and robust segmentation results and reliable uncertainty estimates simultaneously. Our method improved the mean squared error (MSE) loss used for the simple natural image classification \cite{sensoyEvidentialDeepLearning2018}, making it more suitable for semantic segmentation of medical images. The main contributions of our work can be summarized as follows:

\begin{itemize}
    \item An EDL framework was adopted for accurate brain tumor segmentation, which can quantify the uncertainty of segmentation results and improve the reliability and robustness of segmentation.
    \item A novel training loss was developed based on minimizing the region-based prediction error under the Dirichlet prior distribution. Theoretical properties are fully provided to guarantee the evidential learning of model.
    \item A new evaluation metric called soft uncertainty-error overlap (sUEO) was designed for uncertainty estimation to assess the model’s ability to localize segmentation errors more easily.
    \item The robustness of the segmentation accuracy and uncertainty quantification of the proposed method is comparatively evaluated on the BraTS2020 dataset. The effectiveness and efficiency of the novel loss function were verified.
\end{itemize}

The rest of the paper is structured as follows: Section \ref{s2} briefly introduces EDL and recent development. Section \ref{s3} details our segmentation framework, including EDL and novel loss functions. Section \ref{s4} illustrates the experimental setup and evaluation metrics, and the results are analyzed and discussed in Section \ref{s5}. The conclusion and future research directions are given in Section \ref{s6}.

\section{Related Work}\label{s2}
Despite many uncertainty estimation methods mentioned, our proposed framework resorts to arguably the most cutting-edge methodology, EDL, for this purpose. The rest of the section presents the principles of EDL (Subsection \ref{s21}) and a brief overview (Subsection \ref{s22}) of the scarce contributions in which EDL has been utilized for tumor segmentation.

\subsection{Principles of EDL}\label{s21}
Evidence Deep Learning (EDL) is based on Dempster-Shafer Evidence Theory (DST) \cite{dempsterGeneralizationBayesianInference2008}, which is a generalization of Bayesian theory to subjective probability. It assigns belief masses to subsets of a discerning frame, representing a set of exclusive potential states, such as possible class labels for a voxel. A belief mass can be assigned to any subset of the frame. Assigning all belief masses to the entire frame represents the opinion that the truth can be any possible state, e.g., any label is equally likely.

The belief distribution of DST in the discerning framework can be formalized as a Dirichlet distribution by Subjective Logic (SL) \cite{josangSubjectiveLogic2016}. For a voxel $i$, the Dirichlet distribution $Dir\left(\boldsymbol{\alpha}_{i}\right)$ is parameterized by a vector of Dirichlet parameters $\alpha_{ij}$ for $K$ classes, where $j$ denotes the $j$-th class. (The denotations of subscripts $i$ and $j$ hold for the entire paper.) The neural network collects evidence $e_{ij}$ from the input data, a measure of support that facilitates classifying samples into the class $j$. The belief mass distribution, i.e., subjective opinion, in \cite{sensoyEvidentialDeepLearning2018} corresponds to a Dirichlet distribution with parameter $\alpha_{ij}=e_{ij}+1$.

As a result, it is equivalent to placing a Dirichlet distribution on the predicted categorical distribution, allowing a single network to output different predictions. The output layer of an EDL-based neural network parameterizes a simplex distribution representing the probability distribution of class assignments. The softmax/sigmoid classification layer is replaced with a ReLU activation layer that outputs non-negative continuous values, resulting in $e_{ij}$. The vector of predicted classification probabilities can be computed by:
\begin{equation}
\small
    {\hat{p}}_{ij}=\frac{\ \alpha_{ij}}{S_i}, 
\end{equation}
where $S_i=\sum_{j=1}^{K}\alpha_{ij}$ is called the Dirichlet strength. The class probability vector for voxel i given by $\vb{p}_{i}$ is modeled as a random vector drawn from the Dirichlet distribution \cite{tsiligkaridisInformationAwareMaxnorm2021}.

Let $\vb{y}_{i}$ be the one-hot encoded labels with $y_{ik}=1$ and $y_{ij}=0$ for all $j\neq k$. Treating the Dirichlet distribution $Dir\left(\vb{p}_{i} \mid \boldsymbol{\alpha}_{i}\right)$ as a prior on the multinomial likelihood $Mult(\vb{y}_{i} \mid \vb{p}_{i})$, one can minimize the negative logarithm of the marginal likelihood: 
\begin{align}
\footnotesize
\begin{split}
    \mathcal{L}_{\text{ML},i}&=-\log\left(\int{\prod_{j=1}^{K}{p_{ij}}^{y_{ij}}\frac{1}{\mathcal{B}(\boldsymbol{\alpha}_{i})}\prod_{j=1}^{K}{p_{ij}}^{\alpha_{ij}-1}d\vb{p}_{i}}\right)\\&=\sum_{j=1}^{K}{y_{ij}\left(\log\left(S_i\right)-\log\left(\alpha_{ij}\right)\right)},
\end{split}
\end{align}
where $\mathcal{B}$ is the multinomial beta function \cite{kotzContinuousMultivariateDistributions2005a}. Alternatively, one can minimize the Bayes risk of the cross-entropy loss:
\begin{align}
\footnotesize
\begin{split}
    \mathcal{L}_{\text{CE},i}&=\int{\left[-\sum_{j=1}^{K}{y_{ij}\log(p_{ij})}\right]\frac{1}{\mathcal{B}(\boldsymbol{\alpha}_{i})}\prod_{j=1}^{K}{p_{ij}}^{\alpha_{ij}-1}d\vb{p}_{i}}\nonumber
\end{split}\\
\footnotesize
\begin{split}
    &=\sum_{j=1}^{K}{y_{ij}\left(\psi\left(S_i\right)-\psi\left(\alpha_{ij}\right)\right)},
\end{split}
\end{align}
or the mean squared error:
\begin{align}
\footnotesize
\begin{split}
    \mathcal{L}_{\text{MSE},i}&=\int{\lVert\vb{y}_{i}-\vb{p}_{i}\rVert^2\frac{1}{\mathcal{B}(\boldsymbol{\alpha}_{i})}\prod_{j=1}^{K}{p_{ij}}^{\alpha_{ij}-1}d\vb{p}_{i}}\\&=\sum_{j=1}^{K}{\left(y_{ij}-{\hat{p}}_{ij}\right)^2+\frac{{\hat{p}}_{ij}\left(1-{\hat{p}}_{ij}\right)}{\left(S_i+1\right)}},
\end{split}
\end{align}
where $\psi$ refers to the digamma function \cite{moralesConstructionDigammaFunction2008}. Sensoy et al. \cite{sensoyEvidentialDeepLearning2018} observed that $\mathcal{L}_{\text{ML},i}$ and $\mathcal{L}_{\text{CE},i}$ produced excessively high belief masses and were less stable than $\mathcal{L}_{\text{MSE},i}$. This can be attributed to the fact that these two loss functions encourage maximizing the correct likelihood.

\subsection{Related Work of EDL}\label{s22}
Sensoy et al. \cite{sensoyEvidentialDeepLearning2018} used the MSE loss for natural image classification. They showed that the loss decreases as the correct class parameter grows and decreases when the largest incorrect parameter decays. Furthermore, they integrated the KL divergence loss to narrow the error class parameters further. However, the properties of the aggregated loss function were not shown, and the behavior of the loss was not studied for all parameters. Also, for the image classification problem, \cite{tsiligkaridisInformationAwareMaxnorm2021} improved the square-norm of MSE loss to max-norm and achieved higher performance. Because max-norm minimizes the highest class prediction error, and square-norm minimizes the total sum of squares, which is more susceptible to outliers. However, this situation may not be applicable for tumor segmentation with severe class imbalance. The TBraTS network \cite{zouTBraTSTrustedBrain2022} attempted to apply EDL's CE loss to brain tumor segmentation. In order to improve the segmentation accuracy, the network output was additionally passed through the softmax layer to calculate the soft Dice loss, which was added with the CE loss. However, this increases training costs and complexity, and an incomplete deployment of the EDL framework may cause the network to fail to produce true evidence values. Different from these methods that employed MSE or CE loss, our method minimized region-based prediction error (soft Dice loss) under the Dirichlet prior distribution to facilitate tumor segmentation performance of the EDL framework. 

\section{Method}\label{s3}

This section details our approach, a novel region-based EDL framework for 3D brain tumor segmentation (Subsection \ref{s31}) and describes how we quantify the uncertainty (Subsection \ref{s32}).

\subsection{Region-Based Evidential Deep Learning}\label{s31}

For semantic segmentation of medical images, it is important to consider the accuracy of segmented regions in addition to standard classification errors. Hence, we proposed a region-based objective to minimize the expected prediction error in the EDL framework while maintaining high segmentation accuracy. Unlike Zou et al. \cite{zouTBraTSTrustedBrain2022} who added the soft Dice (sDice) loss based on the result of softmax activation to $\mathcal{L}_{\text{CE},i}$, we directly minimized the Bayes risk of sDice loss:
\begin{align}
\footnotesize
\begin{split}
    \text{sDice}&=\frac{1}{K}\sum_{j=1}^{K}{1-\frac{2\sum_{i}{y_{ij}p_{ij}}}{\sum_{i}{{y_{ij}}^2+{p_{ij}}^2}}},
\end{split}\\
\footnotesize
\begin{split}
    \mathcal{L}_{\text{DICE}}&=\int{\left[\text{sDice}\right]\frac{1}{\mathcal{B}(\boldsymbol{\alpha}_{i})}\prod_{j=1}^{K}{p_{ij}}^{\alpha_{ij}-1}d\vb{p}_{i}}\\
    &=\frac{1}{K}\sum_{j=1}^{K}\mathbb{E}\left[1-\frac{2\sum_{i}{y_{ij}p_{ij}}}{\sum_{i}{{y_{ij}}^2+{p_{ij}}^2}}\right]\\
    &=1-\frac{2}{K}\sum_{j=1}^{K}\frac{\sum_{i}{y_{ij}\mathbb{E}\left[p_{ij}\right]}}{\sum_{i}{{y_{ij}}^2+\mathbb{E}\left[{p_{ij}}^2\right]}}.
\end{split}
\end{align}
By using the identity:
\begin{equation}
\small
    \mathbb{E}\left[{p_{ij}}^2\right]={\mathbb{E}\left[p_{ij}\right]}^2+\text{Var}(p_{ij}),
\end{equation}
the equation can be formulated in an easily interpretable form:
\begin{align}
\footnotesize
\begin{split}
    \mathcal{L}_{\text{DICE}}&=1-\frac{2}{K}\sum_{j=1}^{K}\frac{\sum_{i}{y_{ij}{\hat{p}}_{ij}}}{\sum_{i}{{\underbrace{{y_{ij}}^2+{\hat{p}}_{ij}^2}_{\text{sDiceDen}}}+{\underbrace{\frac{{\hat{p}}_{ij}\left(1-{\hat{p}}_{ij}\right)}{\left(S_i+1\right)}}_{\text{Var}}}}}
\end{split}\\
\footnotesize
\begin{split}
    &=1-\frac{2}{K}\sum_{j=1}^{K}\frac{\sum_{i}{y_{ij}\frac{\ \alpha_{ij}}{S_i}}}{\sum_{i}{{y_{ij}}^2+\left(\frac{\ \alpha_{ij}}{S_i}\right)^2+\frac{\alpha_{ij}\left(S_i-\alpha_{ij}\right)}{{S_i}^2\left(S_i+1\right)}}}.\nonumber
\end{split}
\end{align}
By factoring out the denominator of sDice (sDiceDen) and variance (Var), the loss aims to achieve the joint goal of minimizing the region-based prediction error and variance of the Dirichlet experiments generated by the neural network for the training set. 

In order to ensure an effective EDL framework that allows the network to learn to generate subjective opinions from evidence correctly, the loss function needs to have the following properties.

\begin{hyp}
    When the network optimizes, the loss function prioritizes data fitting over variance estimation.
\end{hyp}
\begin{hyp}
    The loss function has a tendency to fit the data.
\end{hyp}
\begin{hyp}
    The loss function avoids generating evidence for all observations it cannot explain.
\end{hyp}

These properties of the proposed DICE loss ($\mathcal{L}_{\text{DICE}}$) can be guaranteed by the following theorems, each numbered one-to-one with the hypothesis. The proofs of all Theorems are presented in the Appendix \ref{appA}.

\begin{theorem}\label{thm1}
    For any $\alpha_{ij}\geq1$, the inequality $sDiceDen > Var$ is satisfied.
\end{theorem}
\begin{theorem}\label{thm2}
    For a given voxel p with the correct label $c$, $\mathcal{L}_{\text{DICE}}$ decreases when new evidence is added to $\alpha_{pc}$ and increases when evidence is removed from $\alpha_{pc}$.
\end{theorem}
\begin{theorem}\label{thm3}
    For a given voxel p with the correct label $c$, $\mathcal{L}_{\text{DICE}}$ decreases when evidence is removed from all incorrect Dirichlet parameters $\alpha_{pw}$ for all $w\neq c$.
\end{theorem}

To summarize, Theorems \ref{thm1} to \ref{thm3} demonstrate that the proposed loss function can optimize the neural network to provide more evidence for the correct class of each voxel while avoiding misclassification by discarding misleading evidence. By accumulating evidence, the loss also tends to reduce the variance of its predictions on the training set, but only if the additional evidence leads to a better fit to the data.

Furthermore, to further minimize the contribution of parameters associated with incorrect classes, a KL divergence loss function is introduced to shrink their evidence to 0 as follows:
\begin{align}
\footnotesize
\begin{split}
    \mathcal{L}_{\text{KL},i}=&\log\left(\frac{\Gamma\left(\sum_{j=1}^{K}{\widetilde{\alpha}}_{ij}\right)}{\Gamma\left(K\right)\prod_{j=1}^{K}\Gamma\left({\widetilde{\alpha}}_{ij}\right)}\right)\\&+\sum_{j=1}^{K}\left({\widetilde{\alpha}}_{ij}-1\right)\left[\psi\left({\widetilde{\alpha}}_{ij}\right)-\psi\left(\sum_{j=1}^{K}{\widetilde{\alpha}}_{ij}\right)\right],
\end{split}
\end{align}
where $\Gamma(\cdot)$ is the gamma function \cite{moralesConstructionDigammaFunction2008} and ${\widetilde{\boldsymbol{\alpha}}}_{i}=\vb{y}_{i}+\left(\mathbf{1}-\vb{y}_{i}\right)\bigodot\boldsymbol{\alpha}_{i}$ is the Dirichlet parameters after removal of the non-misleading evidence. The following theorem shows a desirable monotonicity property of this regularization loss as a supplementary to \cite{sensoyEvidentialDeepLearning2018}.

\begin{theorem}\label{thm4}
    For a voxel $i$ with the correct label $c$, $\mathcal{L}_{\text{KL},i}$ increases in $\alpha_{iw}$ for all $w\neq c$.
\end{theorem}

Theorems \ref{thm3} and \ref{thm4} show that the strength of parameters associated with misleading results is expected to decrease during training. Since the parameters are all expected to be minimized, the preferred behavior of the proposed loss function results in a higher uncertainty of misclassification.

The final loss function is defined as:
\begin{equation}
\small
    \mathcal{L}_{\text{EDL}}=\mathcal{L}_{\text{DICE}}+\lambda\mathcal{L}_{\text{KL,mean}},
\end{equation}
where $\mathcal{L}_{\text{KL,mean}}$ is the mean KL divergence loss over all voxels and $\lambda$ is an annealing coefficient. The KL divergence loss is gradually introduced by $\lambda$ for a stable training due to its strong regularization effect. The annealing scheme is set to reach a maximum $\frac{1}{10}$ as:
$\lambda=\frac{1}{10}{\min\left(1,\ \frac{n}{100}\right)}^2$
where $n$ is the current epoch. 

In addition, the weighted sDice loss, $\mathcal{L}_{\text{wDICE}}$, is also proposed to ease the class imbalance between tumor and background voxels. The weight for each segmentation class is one minus the ratio of foreground voxels to background voxels. Since the weights are all positive and class-wise, all theoretical properties of the loss function still hold.

Furthermore, the parameter of Dirichlet distribution in our framework is re-defined as:
\begin{equation}
\small
    \alpha_{ij}=\left(e_{ij}+1\right)^2.
\end{equation}
Unlike \cite{sensoyEvidentialDeepLearning2018} defined the Dirichlet $\alpha_{ij}=e_{ij}+1$, the alternative formula allows the network to output high Dirichlet parameters more easily. This avoids the defect that it is almost impossible for the network to express a high degree of uncertainty for a particular outcome since each outcome gives a minimal proof of one, i.e., $\alpha_{ij}\geq1$.

\subsection{Uncertainty Quantification}\label{s32}

Calculating the predictive entropy (PE) is a common way to quantify uncertainty. Based on the information theory, PE uses confidence scores of predictions to calculate the total uncertainty for a voxel i, which is defined as:
\begin{equation}
\small
    \mathcal{H}(\vb{p}_{i})=-\sum_{j=1}^{K}{p_{ij}\log(p_{ij})},
\end{equation}
where $\vb{p}_{i}$ is the confidence score vector \cite{malininReverseKLDivergenceTraining2019}. In order to better compare different methods, we normalized the PE by its maximum possible value as:
\begin{equation}
\small
    \mathcal{H}(\vb{p}_{i})=-\frac{1}{\log(K)}\sum_{j=1}^{K}{p_{ij}\log(p_{ij})}.
\end{equation}
As a result, the value range of normalized predictive entropy (NPE) is normalized to $[0, 1]$, where 1 implies the maximum uncertainty and 0 implies the absolutely confident prediction.

\section{Experiment Setup}\label{s4}

Experiments on the standard benchmark (BraTS 2020) were conducted to compare different techniques for uncertainty quantification and evaluate qualitatively the produced segmentation along with the uncertainty associated with each voxel. We first present the implementation details (Subsection \ref{s41}) and then introduce the models (Subsection \ref{s42}) and evaluation metrics (Subsection \ref{s43}) for comparative experiments.

\subsection{Data Acquisition and Processing}\label{s41}
The BraTS 2020 \cite{bakasAdvancingCancerGenome2017,bakasIdentifyingBestMachine2019,menzeMultimodalBrainTumor2015} dataset comprises brain MRI images of various scanners and protocols. The ground truth (GT) label includes the GD-enhancing tumor (ET), peritumoral edema (ED), and necrotic and non-enhancing tumor core (NCR + NET). The segmentation masks were evaluated on three tumor subregions: the ET, the tumor core (TC = ET + NCR + NET), and the whole tumor (WT = ET + NCR + NET + ED). Four MRI modalities of T1, T1ce, T2, and T2-FLAIR were co-registered with a size of 240 $\times$ 240 $\times$ 155. They were then interpolated to $1 mm^3$ and skull-stripped. Since GT labels are only available for the training set (369 cases), we split the original training set into a new training set of 236 cases, a validation set of 59 cases, and a test set of 74 cases. 

All images are cropped to 160 $\times$ 192 $\times$ 128 to reduce computational waste in the background and are then preprocessed by intensity normalization. During the training, various data augmentation techniques were applied on-the-fly as in \cite{liLKAUNet3DLargeKernel2022} to artificially increase the dataset size and minimize the risk of overfitting.

\subsection{Model Training and Optimization} \label{s42}

\begin{figure*}[h]
\includegraphics[width=\textwidth]{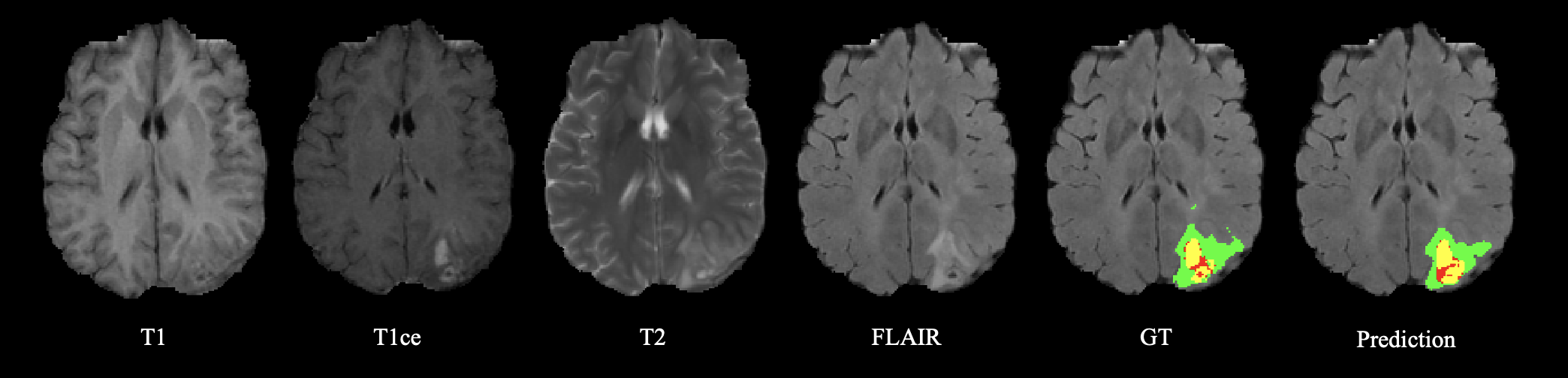}
\caption{Representative visual segmentation results of the proposed region-based EDL method on the BraTS 2020 test set. The labels are enhancing tumor (yellow), edema (green), and necrotic and non-enhancing tumor (red).} 
\label{fig1}
\end{figure*}

We chose LKAU-Net \cite{liLKAUNet3DLargeKernel2022,liLargeKernelAttention3D2022} as our Base network model. All softmax/sigmoid layers in the Base network were replaced with ReLU activation layers as described in the previous section. For comparison, we used different loss functions to optimize the network: $\mathcal{L}_{\text{CE},i}$, $\mathcal{L}_{\text{MSE},i}$, $\mathcal{L}_{\text{DICE}}$, and $\mathcal{L}_{\text{wDICE}}$. Since the evaluation would be based on more meaningful tumor subregions, the network was trained to segment each overlapping subregion separately. However, we also trained the network for multi-class segmentation of the basic labels using $\mathcal{L}_{\text{DICE}}$ of $K=4$ for ablation study.

In addition, we also employed training strategies of Ensemble \cite{lakshminarayananSimpleScalablePredictive2017}, Dropout \cite{galDropoutBayesianApproximation2016}, and TBraTS \cite{zouTBraTSTrustedBrain2022}. For Ensemble, we trained five networks with different initialized weights, which has proven to be sufficient in practice \cite{ovadiaCanYouTrust2019}. Dropout layers (factor of 0.5) were added to the deepest three layers of the Base network to handle high-level features, which is the most efficient \cite{kendallBayesianSegNetModel2017}.These layers were also active during inference, and the same images were passed 10 times to quantify the prediction uncertainty \cite{galDropoutBayesianApproximation2016}. Previous research has found that a sampling rate of 10 is adequate for reasonable uncertainty estimation \cite{galDropoutBayesianApproximation2016}. Moreover, we used the strategy of TBraTS, which combined existing losses for multi-label segmentation.

The adaptive moment estimator (Adam) optimizer was used to optimize all networks in 200 epochs, with a batch size of 1 and an initial learning rate of 0.0003. Experiments were implemented using PyTorch 1.10 on NVIDIA GeForce RTX 3090 GPUs.

\subsection{Evaluation Metrics} \label{s43}
Our method was evaluated using the independent test set of BraTS 2020 (74 cases). The segmentation performance was evaluated using the Dice score, which is defined as:
\begin{equation}
\small
    \text{Dice} = \frac{2\lvert \mathcal{X} \cap \mathcal{Y} \rvert}{\lvert \mathcal{X} \rvert+\lvert \mathcal{Y}\rvert},
\end{equation}
where $\mathcal{X}$ and $\mathcal{Y}$ are sets of GT and prediction. The Dice score measures spatial overlap between the GT and segmentation results, where a score of 1 indicate a complete overlap. 

In addition, the following metrics were utilized to evaluate uncertainty estimation: expected calibration error (ECE) \cite{jungoAnalyzingQualityChallenges2020}, soft uncertainty-error overlap (sUEO), and BraTS score (BraS) \cite{mehtaQUBraTSMICCAIBraTS2021}. ECE is defined by the absolute calibration error between the confidence interval and the accuracy interval ($c_m$ and $a_m$, where $m$ is the $m$-th bin defined in the interval $[0, 1]$), weighted by the number of voxels ($n_m$) in the interval. ECE is given by
\begin{equation}
\small
    \text{ECE}=\sum_{m=1}^{M}\frac{n_m}{N}\lvert c_m-a_m \rvert,
\end{equation}
where N and M are the total numbers of voxels and bins, and the confidence is calculated by one minus the uncertainty. ECE ranges from 0 to 1, where lower values indicate better calibration. To reduce the effect of the large, confident, and accurate extracranial regions typically found in brain tumor MRI, we only considered voxels within the brain. Improved on the uncertainty-error overlap (UEO) \cite{jungoAnalyzingQualityChallenges2020}, we proposed the soft uncertainty-error overlap (sUEO) that directly uses the uncertainty quantities ($u_i$) to measure the overlap:
\begin{equation}
\small
    \text{sUEO}=\frac{2\sum_{i}{y_iu_i}}{\sum_{i}{{y_i}^2+{u_i}^2}}.
\end{equation}
sUEO does not require thresholding the uncertainty map, which can save time optimizing the threshold over the validation set. It shows whether a model can precisely localize segmentation errors. Moreover, the comprehensive BraS is defined by:
\begin{align}
\small
\begin{split}
     \text{BraS}=&\frac{1}{3}\left[\text{{AUC}}_\text{{Dice}}+(1-\text{{AUC}}_\text{{FTP}}) \right.\\
     &\left.+(1-\text{{AUC}}_\text{{FTN}})\right],
\end{split}
\end{align}
where $\text{{AUC}}_\text{{Dice}}$, $\text{{AUC}}_\text{{FTP}}$, and $\text{{AUC}}_\text{{FTN}}$ are area under three curves: 1) Dice vs. confidence threshold, 2) ratio of filtered True Positives (FTP) vs. confidence threshold, and 3) ratio of filtered True Negatives (FTN) vs. confidence threshold. The curves are plotted against the segmentation filtered by different confidence levels, which only voxels with confidence greater than the threshold retain. This metric rewards uncertainty estimates that yield high confidence for correct segmentations or assigns a low confidence level to incorrect segmentations and penalizes uncertainty measures that result in a higher percentage of under-confident correct segmentations.

\section{Results and Discussion}\label{s5}

This section first evaluates the performance of the novel region-based EDL framework for brain tumor segmentation and uncertainty quantification through experiments on the original dataset (Subsection \ref{s51}). It then examines its robustness by applying various image processing techniques to the test image data (Subsection \ref{s52}).

\subsection{Segmentation and Uncertainty Estimation}\label{s51}

Our method generated comparable segmentation results with the GT labels, as visualized in Figure \ref{fig1}. The quantitative results of our methods averaged over three tumor subregions were compared in Table \ref{tab1}. Although the proposed DICE and wDICE loss functions achieved the highest Dice scores (0.791 and 0.793) among all EDL-based methods, Ensemble and Dropout methods performed slightly more accurately in segmentation (0.807 and 0.804). The success of the Ensemble was attributed to the variance reduction by combining predictions prone to various errors. However, the dominance of the proposed region-based losses in all EDL frameworks still proved their effectiveness in improving segmentation performance. Compared to CE-based or MSE-based losses, the DICE-based losses remarkably improved the Dice score by 0.01. 

\begin{figure*}[!h]
\centering
\includegraphics[width=0.75\textwidth]{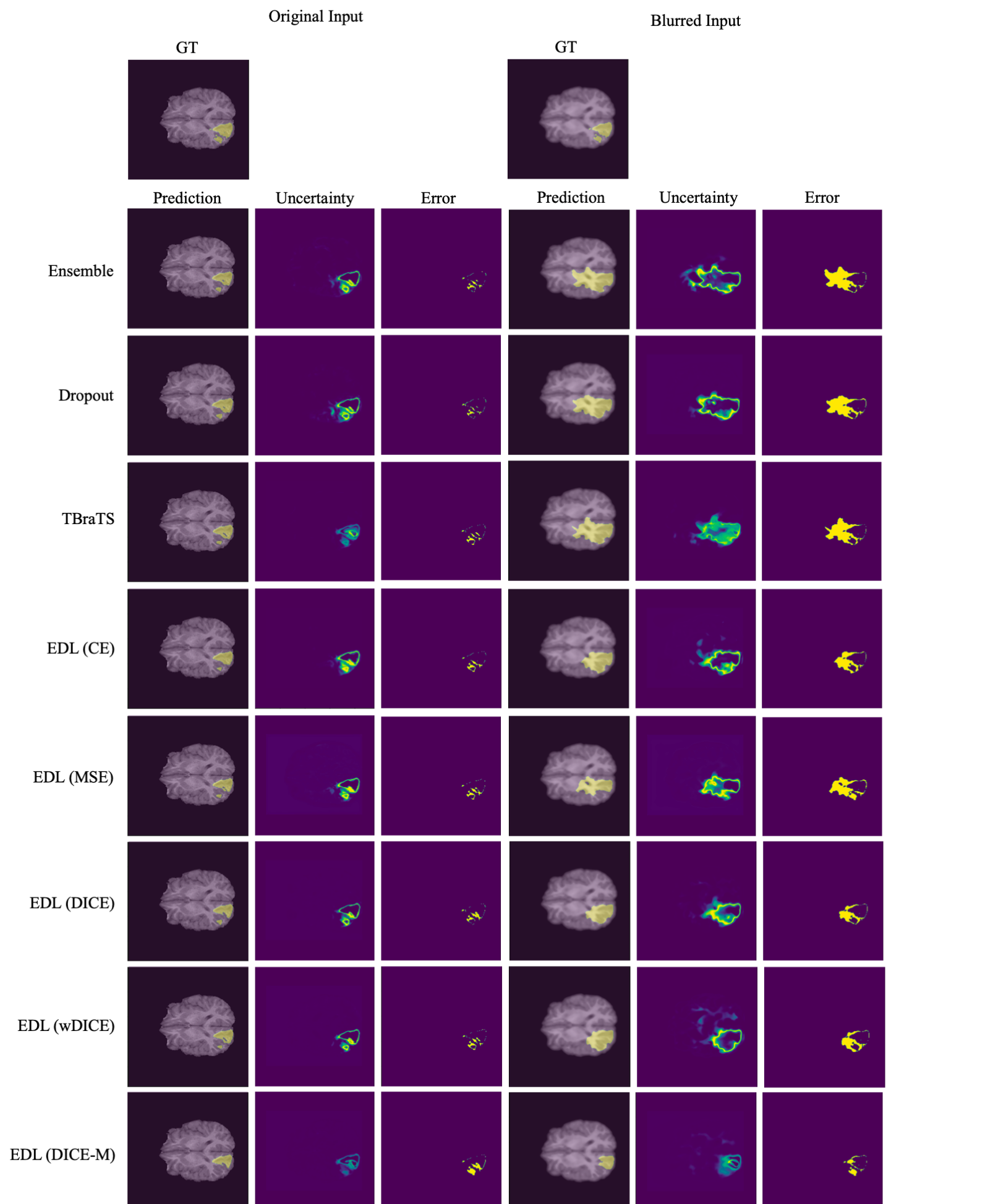}
\caption{Representative visual results of the whole tumor (WT) produced by different uncertainty estimation methods on the BraTS 2020 test set. The right half of the figure was evaluated on the test images blurred by a Gaussian filter of sigma = 1.5.} 
\label{fig2}
\end{figure*}

As for the uncertainty estimation, Ensemble and Dropout still showed their advantages in accuracy. They obtained the lowest ECE metrics of 0.009 and 0,010, which indicated they were well-calibrated. However, our methods achieved the highest sUEO and BraS of 0.420 and 0.875, showing their ability to precisely locate errors and assign uncertainty. The EDL model optimized by the proposed wDice loss generated the most accurate uncertainty map to indicate the potential false predictions. On the other hand, the proposed EDL (DICE) model made the most reliable uncertainty estimation, maintaining the lowest error while thresholding along the uncertainty dimension. The advantages of our region-based EDL methods are also shown in Figure \ref{fig2}. The proposed methods had the most precise uncertainty map consistent with the error map. Ideally, a learned model only give high uncertainty for a possible erroneous prediction. Despite the high segmentation accuracy, both Ensemble and Dropout methods generated more uncertainty around mask boundaries and other correct regions, leading to inferior uncertainty estimation performance in terms of sUEO and BraS. It is also worth mentioning that EDL models trained to segment each tumor subregion separately outperformed the ones trained with multi-class labels (DICE-M).

\begin{table}[!h]
\centering
\caption{Quantitative comparisons of different uncertainty estimation methods on the BraTS 2020 test set. (Bold numbers: best results)}
\label{tab1}
\begin{tabular}{@{}lcccc@{}}
\toprule
\textbf{Method} & \multicolumn{1}{c}{\textbf{Dice$\uparrow$}} & \multicolumn{1}{c}{\textbf{ECE$\downarrow$}} & \multicolumn{1}{c}{\textbf{sUEO$\uparrow$}} & \multicolumn{1}{c}{\textbf{BraS$\uparrow$}} \\ \midrule
Ensemble & \textbf{0.807} & 0.010 & 0.409 & 0.873 \\
Dropout & 0.804 & \textbf{0.009} & 0.412 & 0.869 \\
TBraTS & 0.790 & 0.019 & 0.383 & 0.862 \\
EDL (CE) & 0.783 & 0.038 & 0.325 & 0.829 \\
EDL (MSE) & 0.783 & 0.038 & 0.325 & 0.829 \\
EDL (DICE) & 0.791 & 0.017 & 0.414 & \textbf{0.875} \\
EDL (wDICE) & 0.793 & 0.016 & \textbf{0.420} & 0.874 \\
EDL (DICE-M) & 0.771 & 0.035 & 0.283 & 0.860 \\ \bottomrule
\end{tabular}
\end{table}

In addition, the inference runtimes of the uncertainty estimation methods on one sample are reported in Table \ref{tab2}. The runtime of all EDL-based methods is lower than the others. This is because both Ensemble and Dropout use multiple sampling mechanisms at inference time to obtain uncertainty estimates.

\begin{table}[!h]
\centering
\caption{Inference runtimes of different uncertainty estimation methods for one image.}
\label{tab2}
\begin{tabular}{@{}lr@{}}
\toprule
\textbf{Method} & \textbf{Runtime (s)} \\ \midrule
Ensemble & 6.94 ± 0.05 \\
Dropout & 63.68 ± 0.17 \\
TBraTS & 3.22 ± 0.05 \\
EDL (CE) & 3.39 ± 0.06 \\
EDL (MSE) & 3.41 ± 0.05 \\
EDL (DICE) & 3.38 ± 0.03 \\
EDL (wDICE) & 3.40 ± 0.02 \\
EDL (DICE-M) & 3.23 ± 0.04 \\ \bottomrule
\end{tabular}
\end{table}

\subsection{Robustness Experiment}\label{s52}

\begin{table*}[!h]
\setlength{\tabcolsep}{1.5pt}
\footnotesize
\centering
\caption{Quantitative comparisons of different uncertainty estimation methods on preprocessed BraTS 2020 test set. (Bold numbers: best results)}
\label{tab3}
\begin{tabular}{@{}lcccccccccccccc@{}}
\toprule
\multirow{3}{*}{\textbf{Method}} & \multicolumn{4}{l}{\textbf{Blurred}} && \multicolumn{4}{l}{\textbf{Noisy}} && \multicolumn{4}{l}{\textbf{Gamma Corrected}} \\
 & \multicolumn{1}{l}{\textbf{Dice}} & \multicolumn{1}{l}{\textbf{ECE}} & \multicolumn{1}{l}{\textbf{sUEO}} & \multicolumn{1}{l}{\textbf{BraS}} && \multicolumn{1}{l}{\textbf{Dice}} & \multicolumn{1}{l}{\textbf{ECE}} & \multicolumn{1}{l}{\textbf{sUEO}} & \multicolumn{1}{l}{\textbf{BraS}} && \multicolumn{1}{l}{\textbf{Dice}} & \multicolumn{1}{l}{\textbf{ECE}} & \multicolumn{1}{l}{\textbf{sUEO}} & \multicolumn{1}{l}{\textbf{BraS}} \\
 & \multicolumn{1}{l}{$\uparrow$} & \multicolumn{1}{l}{$\downarrow$} & \multicolumn{1}{l}{$\uparrow$} & \multicolumn{1}{l}{$\uparrow$} && \multicolumn{1}{l}{$\uparrow$} & \multicolumn{1}{l}{$\downarrow$} & \multicolumn{1}{l}{$\uparrow$} & \multicolumn{1}{l}{$\uparrow$} && \multicolumn{1}{l}{$\uparrow$} & \multicolumn{1}{l}{$\downarrow$} & \multicolumn{1}{l}{$\uparrow$} & \multicolumn{1}{l}{$\uparrow$} \\ \midrule
Ensemble & 0.561 & 0.025 & 0.408 & 0.772 && 0.751 & 0.024 & 0.407 & 0.780 && 0.702 & 0.035 & 0.388 & 0.730 \\
Dropout & 0.544 & 0.026 & 0.401 & 0.772 && 0.729 & 0.021 & 0.405 & 0.774 && 0.693 & 0.040 & 0.392 & 0.724 \\
TBraTS & 0.546 & 0.025 & 0.384 & 0.636 && 0.756 & 0.024 & 0.387 & 0.654 && 0.722 & 0.044 & 0.345 & 0.680 \\
EDL (CE) & 0.562 & 0.028 & 0.416 & 0.775 && 0.756 & 0.022 & 0.440 & 0.785 && 0.659 & 0.053 & 0.460 & 0.751 \\
EDL (MSE) & 0.559 & 0.038 & 0.419 & 0.770 && 0.739 & 0.040 & 0.391 & 0.773 && 0.632 & 0.071 & 0.413 & 0.700 \\
EDL (DICE) & 0.571 & \textbf{0.024} & \textbf{0.458} & \textbf{0.796} && 0.769 & \textbf{0.022} & 0.447 & \textbf{0.803} && \textbf{0.711} & \textbf{0.033} & \textbf{0.480} & \textbf{0.768} \\
EDL (wDICE) & \textbf{0.572} & 0.025 & 0.442 & 0.793 && \textbf{0.771} & 0.023 & \textbf{0.449} & 0.799 && 0.707 & 0.036 & 0.448 & 0.763 \\
EDL (DICE-M) & 0.438 & 0.037 & 0.343 & 0.682 && 0.758 & 0.027 & 0.398 & 0.723 && 0.704 & 0.037 & 0.350 & 0.626 \\ \bottomrule
\end{tabular}
\end{table*}

\begin{figure*}[!h]
\centering
\includegraphics[width=0.75\textwidth]{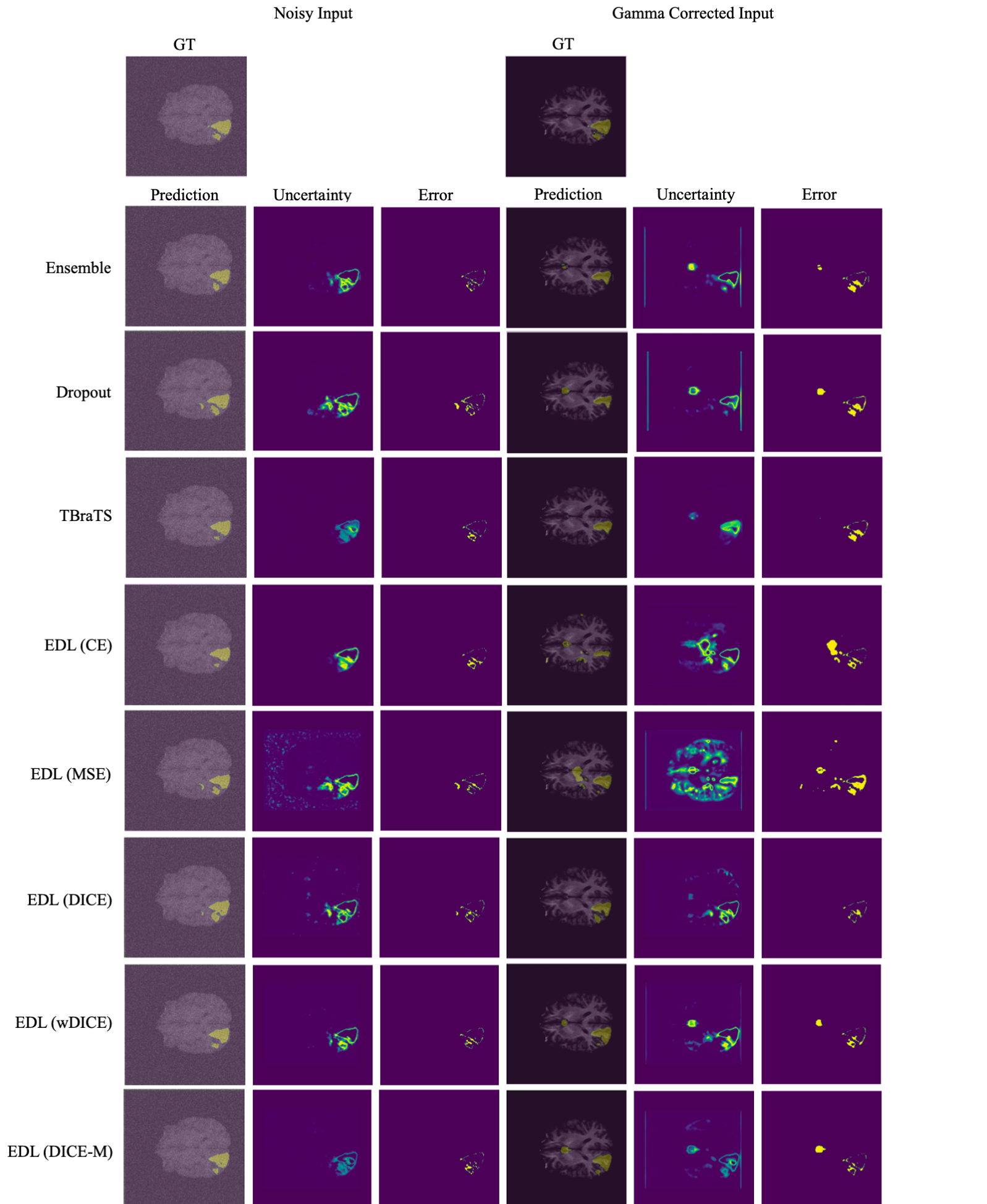}
\caption{Representative visual results of the whole tumor (WT) produced by different uncertainty estimation methods on the BraTS 2020 test set. The left half of the figure was evaluated on the test images added with Gaussian noise of variance = 1.5. The right half was evaluated on the test images after Gamma correction of gamma = 5.} 
\label{fig3}
\end{figure*}

To verify the robustness of the segmentation model, we applied several image processing techniques to simulate the low-quality acquisition that usually happens in actual practice. We first blurred the four modalities of the MRI images using a Gaussian filter with sigma = 1.5. Subsequently, we re-evaluated the performance of all methods for segmentation and uncertainty estimation, as shown in Table \ref{tab3}. We can observe that with the addition of Gaussian blur, the segmentation performance of all methods dropped significantly, especially Ensemble and Dropout. Our method leaped to the highest Dice metric of 0.572 for blurry images, demonstrating its robustness. By comparing the segmentation results with the original input and high-noise input in Figure \ref{fig2}, it can be seen that the EDL using our loss function segmented the WT region more accurately than all other methods. This is due to the evidence extracted from the data that produced these subjective opinions.

Furthermore, our method exhibited the most reliable uncertainty quantification on blurred images compared to other uncertainty estimation methods. As shown in Table \ref{tab3}, all uncertainty evaluation metrics decreased, except for sUEO for all EDL-based methods. This might benefit from the robust evidence captured by the EDL segmentation framework, where the advantage is more noticeable with larger error regions. The proposed EDL (DICE) method achieved top performance, especially for generating reliable uncertainty maps. Besides showing the robustness of our method, this again demonstrated the importance of region-based loss for locating semantic segmentation errors. As shown in the right half of Figure \ref{fig2}, the uncertainty map generated by the EDL (DICE) method is the most relevant to the error map. Unlike other methods that only generate high uncertainty at the edges of the predicted mask, the proposed method can indicate potential error regions inside masks. It is true that regional boundaries should a priori undergo higher uncertainty, but this high uncertainty should not be assumed to be exclusive of these regions. This is what is favored by our proposed methods, as they declare high uncertainty in delimiting and non-delimiting regions of the segmented image. This demonstrates the potential of our proposed method for clinical application. Potential error regions fed back by the model can assist in automatic correction or quality quantification of tumor segmentation.

To enrich the robustness experiment, we further applied Gaussian noise and Gamma correction to the original input to simulate the noise and the contrast variability introduced by imaging or enhancement technique. With the Gaussian noise of variance = 1.5, the segmentation performance of all methods was superior to burred ones, as shown in Table \ref{tab3}. The proposed EDL (wDICE) remained the top in The proposed EDL (wDICE) method remained the top in terms of the Dice metric (0.771), followed by the proposed EDL (DICE) with Dice 0.769. The metrics for uncertainty estimation resembled blurred input, while the sUEO of EDL (wDICE) became 0.002 higher than that of EDL (DICE). However, as for the result after Gamma correction with gamma = 5, the proposed EDL (DICE) method excelled in all metrics, which showed its robustness to unexpected contrast variance. 

These observations can also be visually inspected in Figure \ref{fig3}. Compared to other methods, our methods provided more precise uncertainty maps. Ensemble and Dropout models had trouble handling the boundaries, especially for Gamma corrected input. The extracranial area is no longer zero after Gamma correction, which might cause problems when applying zero-padding. Moreover, their overconfident prediction using softmax/sigmoid is shown for Gamma corrected input where the main error regions were not indicated in the uncertainty map. Non-region-based EDL methods also showed inaccurate uncertainty maps. The shortcoming of using MSE loss in EDL to quantify the uncertainty of medical image segmentation can be seen in Figure 3, which was significantly biased by the interference.

\section{Conclusions and Future Research Directions}\label{s6}

In this paper, we proposed a region-based EDL framework to segment brain tumors and quantify their uncertainty reliably and robustly. We demonstrated that the proposed region-based loss could generate reliable prediction confidence by gathering evidence in the output image by demonstrating four theoretical properties. Our method produced voxel-level uncertainty maps for brain tumor segmentation, which provided additional information on segmentation confidence for cancer diagnosis. Extensive experiments showed that the proposed method is more robust than previous methods on the BraTS 2020 dataset and achieves the best performance in segmentation uncertainty estimation. Furthermore, the novel framework maintained the low computational cost properties of EDL and can be easily integrated into any neural network. 

However, our method was currently inferior to Ensemblmultie and Dropout methods in the segmentation performance of raw images. EDL frameworks with higher segmentation accuracy are worthy of further study. In addition, since the predictive uncertainty can be separated into epistemic and aleatoric uncertainty, future work can also focus on the inherent value for automated diagnosis that uncertainty estimation brings when differentiating between the two sources of uncertainty. The third direction is validating this framework in other diagnostic applications, possibly favoring the fusion of more multimodal information sources. Then, we can assess whether the fusion of different information modes permits a decrease in the overall uncertainty of the model estimated by our EDL segmentation framework.

\backmatter
\small
\bmhead{Acknowledgments}

This study was supported in part by the BHF (TG/18/5/34111, PG/16/78/32402), the ERC IMI (101005122), the H2020 (952172), the MRC (MC/PC/21013), the Royal Society (IEC/NSFC/211235), the Imperial College Undergraduate Research Opportunities Programme (UROP), the NVIDIA Academic Hardware Grant Program, the SABER project supported by Boehringer Ingelheim Ltd, NIHR Imperial Biomedical Research Centre (RDA01), and the UKRI Future Leaders Fellowship (MR/V023799/1). J. Del Ser acknowledges funding support from the Basque Government through grant number IT1456-22 (MATHMODE), as well as the Spanish Centro para el Desarrollo Tecnologico Industrial (CDTI, Ministry of Science and Innovation) through the ``Red Cervera'' Programme (AI4ES project).

\section*{Declarations}

\subsection*{Declaration of competing interest}
The authors have no relevant financial or non-financial interests to disclose. 

\subsection*{Data availability statement}
The datasets generated during and/or analysed during the current study are available from the corresponding author on reasonable request. 

\subsection*{CRediT authorship contribution statement}
\textbf{Hao Li}: Conceptualization, Methodology, Software, Validation, Formal analysis, Investigation, Writing - original draft. \textbf{Yang Nan}: Conceptualization, Methodology, Writing - review and editing, Supervision. \textbf{J. Del Ser}: Conceptualization, Writing - review and editing, Supervision. \textbf{Guang Yang}: Conceptualization, Methodology, Writing - review and editing, Supervision, Funding acquisition.

\begin{footnotesize}
\bibliography{EDL.bib}
\end{footnotesize}


\newpage
\onecolumn
\begin{appendices}
\section{Proofs of Theorems}\label{appA}
This section provides full proofs for Theorems \ref{thm1} to \ref{thm4}.


\begin{proof}[Proof of Theorem~{\upshape\ref{thm1}}]
Since $\frac{\left(S_i-\alpha_{ij}\right)}{\left(S_i+1\right)}<1$ and $\frac{\alpha_{ij}}{{S_i}^2}\leq\left(\frac{\alpha_{ij}}{S_i}\right)^2$,
\begin{equation}
    \left(\frac{\ \alpha_{ij}}{S_i}\right)^2>\frac{\alpha_{ij}\left(S_i-\alpha_{ij}\right)}{{S_i}^2\left(S_i+1\right)}.
\end{equation}
As $y_{ij}\in{0,\ 1}$,
\begin{equation}
    {y_{ij}}^2+\left(\frac{\alpha_{ij}}{S_i}\right)^2>\frac{\alpha_{ij}\left(S_i-\alpha_{ij}\right)}{{S_i}^2\left(S_i+1\right)}.
\end{equation}
\end{proof}


\begin{proof}[Proof of Theorem~{\upshape\ref{thm2}}]
The loss function becomes:
\begin{align}
\footnotesize
\begin{split}
    \mathcal{L}_{DICE}=&1-\frac{2}{K}\left[\underbrace{\frac{\frac{ \alpha_{pc}+\varepsilon}{S_p+\varepsilon}+\overbrace{\sum_{i\neq p}\frac{\alpha_{ic}}{S_i}}^{L_{n,c}}}{1+\left(\frac{ \alpha_{pc}+\varepsilon}{S_p+\varepsilon}\right)^2+\underbrace{\frac{\left(\alpha_{pc}+\varepsilon\right)\left(S_p-\alpha_{pc}\right)}{\left(S_p+\varepsilon\right)^2\left(S_p+1+\varepsilon\right)}}_{{Var}_{pc}}+\underbrace{\sum_{i\neq p}{1+\left(\frac{ \alpha_{ic}}{S_i}\right)^2+\frac{\alpha_{ic}\left(S_i-\alpha_{ic}\right)}{{S_i}^2\left(S_i+1\right)}}}_{L_{d,c}}}}_{C_c}\right.\\
    &\left.+\underbrace{\sum_{j\neq c}^{K}\frac{\sum_{i\neq p}\frac{ \alpha_{ij}}{S_i}}{\left(\frac{ \alpha_{pj}}{S_p+\varepsilon}\right)^2+\frac{\alpha_{pj}\left(S_p-\alpha_{pj}\right)}{\left(S_p+\varepsilon\right)^2\left(S_p+1+\varepsilon\right)}+\sum_{i}{\left(\frac{ \alpha_{ij}}{S_i}\right)^2+\frac{\alpha_{ij}\left(S_i-\alpha_{ij}\right)}{{S_i}^2\left(S_i+1\right)}}}}_{D_c}\right],
\end{split}
\end{align}
where $\varepsilon$ is the change in $\alpha_{pc}$. For $\varepsilon>0$, $\frac{\ \alpha_{pc}+\varepsilon}{S_p+\varepsilon}>\frac{\ \alpha_{pc}}{S_p}$, so ${\hat{p}}_{pc}=\frac{\ \alpha_{pc}}{S_p}$ increases as $\alpha_{pc}$ increases. 

Taking the partial derivative of $C_c$, we get:
\begin{equation}
    \frac{\partial C_c}{\partial{\hat{p}}_{pc}}=\frac{1+{Var}_{pc}+L_{d,c}-2{\hat{p}}_{pc}L_{n,c}-{\hat{p}}_{pc}^2}{\left(1+{Var}_{pc}+L_{d,c}+{\hat{p}}_{pc}^2\right)^2}.
\end{equation}
The numerator can be organized as:
\begin{equation}
    numerator=\left(1-{\hat{p}}_{pc}^2\right)+\left(L_{d,c}-2{\hat{p}}_{pc}L_{n,c}\right)+{Var}_{pc}.
\end{equation}
Since $0<{\hat{p}}_{pc}<1$, $\frac{L_{n,c}}{L_{d,c}}\le\frac{1}{2}$, ${Var}_{pc}\geq0$, and the denominator is non-negative, 
\begin{equation}
    \frac{\partial C_c}{\partial{\hat{p}}_{pc}}>0.
\end{equation}
Thus, $C_c$ increases as ${\hat{p}}_{pc}$ increases, equally as $\alpha_{pc}$ increases.

As for $D_c$, when $\varepsilon>0$, $\left(\frac{\ \alpha_{pj}}{S_p+\varepsilon}\right)^2<\left(\frac{\ \alpha_{pj}}{S_p}\right)^2$ and $\frac{\alpha_{pj}\left(S_p-\alpha_{pj}\right)}{\left(S_p+\varepsilon\right)^2\left(S_p+1+\varepsilon\right)}<\frac{\alpha_{pj}\left(S_p-\alpha_{pj}\right)}{{S_p}^2\left(S_p+1\right)}$. So $D_c$ also increases as $\alpha_{pc}$ increases.

As a result, $\mathcal{L}_{DICE}=1-\frac{2}{K}\left[C_c+D_c\right]$ decreases as $\alpha_{pc}$ increases. The reasoning is the same for $\varepsilon<0$.
\end{proof}


\begin{proof}[Proof of Theorem~{\upshape\ref{thm3}}]
The loss function becomes:
\begin{align}
\footnotesize
\begin{split}
    \mathcal{L}_{DICE}=&1-\frac{2}{K}\left[{\underbrace{\frac{\frac{\ \alpha_{pc}}{S_p+\varepsilon_0}+L_{n,c}}{1+\left(\frac{\ \alpha_{pc}}{S_p+\varepsilon_0}\right)^2+\frac{\alpha_{pc}\left(S_p-\alpha_{pc}\right)}{\left(S_p+\varepsilon_0\right)^2\left(S_p+1+\varepsilon_0\right)}+L_{d,c}}}_{C_w}}\right.\\
    &\left.+{\underbrace{\sum_{j\neq c}^{K}\frac{\overbrace{\sum_{i\neq p}\frac{\ \alpha_{ij}}{S_i}}^{L_{n,j}}}{\left(\frac{\ \alpha_{pj}+\varepsilon_j}{S_p+\varepsilon_j}\right)^2+\frac{\left(\alpha_{pj}+\varepsilon_j\right)\left(S_p-\alpha_{pj}\right)}{\left(S_p+\varepsilon_j\right)^2\left(S_p+1+\varepsilon_j\right)}+{\underbrace{\sum_{i}{\left(\frac{\ \alpha_{ij}}{S_i}\right)^2+\frac{\alpha_{ij}\left(S_i-\alpha_{ij}\right)}{{S_i}^2\left(S_i+1\right)}}}_{L_{d,j}}}}}_{W_w}}\right],
\end{split}
\end{align}
where $\varepsilon_j$ is the change in $\alpha_{pj}$ and $\varepsilon_0=\sum_{j\neq c}^{K}\varepsilon_j$. For $\varepsilon_0<0$ and $j\neq c$, $\frac{\alpha_{pc}}{S_p+\varepsilon_0}>\frac{\alpha_{pc}}{S_p}$, so ${\hat{p}}_{pc}$ increases as $\alpha_{pj}$ increases. 

Taking the partial derivative of $C_w$, we get:
\begin{equation}
    \frac{\partial C_w}{\partial{\hat{p}}_{pc}}=\frac{\left(1-{\hat{p}}_{pc}^2\right)+\left(L_{d,c}-2{\hat{p}}_{pc}L_{n,c}\right)+{Var}_{pc}}{\left(1+{Var}_{pc}+L_{d,c}+{\hat{p}}_{pc}^2\right)^2}.
\end{equation}
Since $0<{\hat{p}}_{pc}<1$, $\frac{L_{n,c}}{L_{d,c}}\le\frac{1}{2}$, ${Var}_{pc}\geq0$, and the denominator is non-negative, 
\begin{equation}
    \frac{\partial C_w}{\partial{\hat{p}}_{pc}}>0.
\end{equation}
Thus, $C_w$ increases as ${\hat{p}}_{pc}$ increases, equally as $\alpha_{pj}$ decreases for all $j\neq c$.

Similarly, when $\varepsilon_j<0$ and $j\neq c$, $\left(\frac{\ \alpha_{pj}+\varepsilon_j}{S_p+\varepsilon_j}\right)^2<\left(\frac{\ \alpha_{pj}}{S_p}\right)^2$, ${\hat{p}}_{pj}=\frac{\ \alpha_{pj}}{S_p}$ decreases as $\alpha_{pj}$ decreases. The partial derivative of $W_w$ is:
\begin{equation}
    \frac{\partial W_w}{\partial{\hat{p}}_{pj}}=\frac{-2{\hat{p}}_{pj}L_{n,j}}{\left(L_{d,j}+{\hat{p}}_{pj}^2\right)^2}.
\end{equation}
Since $0<{\hat{p}}_{pj}<1$, $L_{n,j}>0$, and the denominator is non-negative, 
\begin{equation}
    \frac{\partial W_w}{\partial{\hat{p}}_{pj}}<0.
\end{equation}
Thus, $W_w$ increases as $\alpha_{pj}$ decreases.

As a result, $\mathcal{L}_{DICE}=1-\frac{2}{K}\left[C_w+W_w\right]$ decreases as $\alpha_{pj}$ decreases for all $j\neq c$. The reasoning is the same for $\varepsilon>0$.
\end{proof}

\begin{proof}[Proof of Theorem~{\upshape\ref{thm4}}]
The loss function becomes:
\begin{align}
\begin{split}
    \mathcal{L}_{KL,i}=&{\underbrace{\log\left(\frac{\Gamma\left(\varepsilon_w+\sum_{j=1}^{K}{\widetilde{\alpha}}_{ij}\right)}{\Gamma\left(K\right)\Gamma\left({\widetilde{\alpha}}_{iw}+\varepsilon_w\right)\prod_{j\neq w}^{K}\Gamma\left({\widetilde{\alpha}}_{ij}\right)}\right)}_A}\\
    &+{\underbrace{\left({\widetilde{\alpha}}_{iw}-1+\varepsilon_w\right)\left[\psi\left({\widetilde{\alpha}}_{iw}+\varepsilon_w\right)-\psi\left(\varepsilon_w+\sum_{j=1}^{K}{\widetilde{\alpha}}_{ij}\right)\right]}_B}\\
    &+{\underbrace{\sum_{j=1}^{K}\left({\widetilde{\alpha}}_{ij}-1\right)\left[\psi\left({\widetilde{\alpha}}_{ij}\right)-\psi\left(\varepsilon_w+\sum_{j=1}^{K}{\widetilde{\alpha}}_{ij}\right)\right]}_C},
\end{split}
\end{align}
where $\varepsilon_w$ is the change in ${\widetilde{\alpha}}_{iw}$/$\alpha_{iw}$.

As for $A$, when $\varepsilon_w>0$, 
\begin{equation}
    \frac{\Gamma\left(\varepsilon_w+\sum_{j=1}^{K}{\widetilde{\alpha}}_{ij}\right)}{\Gamma\left({\widetilde{\alpha}}_{iw}+\varepsilon_w\right)}>\frac{\Gamma\left(\sum_{j=1}^{K}{\widetilde{\alpha}}_{ij}\right)}{\Gamma\left({\widetilde{\alpha}}_{iw}\right)},
\end{equation}
since $\sum_{j=1}^{K}{\widetilde{\alpha}}_{ij}>{\widetilde{\alpha}}_{iw}$. Hence, $A$ increases as ${\widetilde{\alpha}}_{iw}$/$\alpha_{iw}$ increases.

Similarly for $B$ and $C$, when $\varepsilon_w>0$, 
\begin{align}
    \psi\left({\widetilde{\alpha}}_{iw}+\varepsilon_w\right)-\psi\left(\varepsilon_w+\sum_{j=1}^{K}{\widetilde{\alpha}}_{ij}\right)&>\psi\left({\widetilde{\alpha}}_{iw}\right)-\psi\left(\sum_{j=1}^{K}{\widetilde{\alpha}}_{ij}\right),\\
    \psi\left({\widetilde{\alpha}}_{ij}\right)-\psi\left(\varepsilon_w+\sum_{j=1}^{K}{\widetilde{\alpha}}_{ij}\right)&>\psi\left({\widetilde{\alpha}}_{ij}\right)-\psi\left(\sum_{j=1}^{K}{\widetilde{\alpha}}_{ij}\right).
\end{align}

As a result, $\mathcal{L}_{KL,i}$ increases as ${\widetilde{\alpha}}_{iw}$/$\alpha_{iw}$ increases.
\end{proof}

\end{appendices}

\end{document}